\documentclass[preprint,nofootinbib]{revtex4}

\usepackage[utf8]{inputenc}
\usepackage{tikz-cd}
\usepackage{cancel}
\usepackage{amsmath,amsthm}
\usepackage{amsfonts}
\usepackage{mathtools}
\usepackage[mathscr]{eucal}
\usepackage[sans]{dsfont}
\usepackage{tensor}
\usepackage{bigints}
\usepackage{calc}
\usepackage{makecell}
\usepackage{bbm}
\usepackage{amsmath}
\usepackage{bm}
\usepackage{verbatim}
\usepackage{hyperref}

\newcommand{\be}{\begin{equation}}
\newcommand{\ee}{\end{equation}}
\newcommand{\beq}{\begin{eqnarray}}
\newcommand{\eeq}{\end{eqnarray}}
\newcommand{\bea}{\begin{eqnarray}}
\newcommand{\eea}{\end{eqnarray}}
\newcommand{\beqn}{\begin{eqnarray}}
\newcommand{\eeqn}{\end{eqnarray}}
\def\bs{\begin{subequations}}
\def\es{\end{subequations}}
\def\a{\alpha}

\def\e{\epsilon}

\def\s{\sigma}

\def\p{\partial}
\newcommand{\Eq}[1]{(\ref{#1})}

\def\rmi{{\rm i} \hspace{.5 mm}}

\def\rmd{{\rm d}}
\def\klm{{k\ell m}}

\def\e{\epsilon}
\newcommand{\rd}{\mathrm{d}}

\def\s{\sigma}

\def\tr{\mathrm{tr}}
\def\nn{\nonumber}

\begin{document}
\title{An Inner Product for 4D Quantum Gravity and the Chern-Simons-Kodama State}

\author{Stephon Alexander}
\email{stephon\_alexander@brown.edu}
\affiliation{Brown Theoretical Physics Center
Department of Physics, Brown University,
Providence, RI 02912, USA}

\author{Laurent Freidel}
\email{lfreidel@perimeterinstitute.ca}
\affiliation{Perimeter Institute for Theoretical Physics,
31 Caroline Street North, Waterloo, Ontario, Canada N2L 2Y5}

\author{Gabriel Herczeg\footnote{Corresponding author}}
\email{gabriel\_herczeg@brown.edu
}
\affiliation{Brown Theoretical Physics Center
Department of Physics, Brown University,
Providence, RI 02912, USA}

\date{\today}

\begin{abstract} 
We demonstrate that reality conditions for the Ashtekar connection imply a non-trivial measure for the inner product of gravitational states in the polarization where the Ashtekar connection is diagonal, and we express the measure as the determinant of a certain first-order differential operator. This result opens the  possibility to perform a non-perturbative analysis of the quantum gravity scalar product.
In this polarization, the Chern-Simons-Kodama state, which solves the constraints of quantum gravity for a certain factor ordering, and which has de Sitter space as a semiclassical limit,  is perturbatively non-normalizable with respect to the na\"{i}ve inner product. Our work reopens the question of whether this state might be 
normalizable when the correct non-perturbative inner product and choice of integration contour are taken into account. As a first step, we perform a semi-classical treatment of the measure by evaluating it on the round three-sphere, viewed as a closed spatial slice of de Sitter. The result is a simple, albeit divergent, infinite product that might serve as a regulator for a more complete treatment of the problem. Additionally, our results suggest deep connections between the problem of computing the norm of the CSK state in quantum gravity and computing the Chern-Simons partition function for a complex group. 
\end{abstract}

\maketitle

\tableofcontents

\section{Introduction}

Only a handful of attempts to unite quantum mechanics with gravity rely solely on consistency with the principles of general relativity and quantum mechanics. Moreover, astrophysical observations 
can inform which theory of quantum gravity is correct, where black holes and the early universe have provided some guidance.  For example, cosmic inflation is a phase of the early universe that is dominated by a constant part of the energy-momentum tensor, which is indistinguishable from a cosmological constant.  Furthermore, our current universe is dominated by a cosmological constant, $\Omega_{\Lambda}$. Both situations suggest the semiclassical state of our past and present universe consists of a de Sitter geometry.  

The paradigm of cosmic inflation relies on quantum 
fluctuations of the metric in de Sitter space to seed
the observed anisotropies in the CMB and the subsequent cosmic large-scale structure.  While the observed amplitudes of these super-horizon quantum fluctuations are perturbative and infrared, they emerge from a trans-Planckian epoch at the beginning of inflation  where quantum effects in gravity are important and non-perturbative \cite{BM}.  The assumption of quantum field theory in a fixed semiclassical spacetime is no longer valid in this setting, and one needs a well-controlled quantum gravitational description of nature to understand the nature of these trans-Planckian degrees of freedom.  

It has been known for some time that the Ashtekar variables provide a gauge-theoretic framework that recovers the full phase space of general relativity.  In these variables, Kodama found an exact solution to the Wheeler-de Witt equation and the other constraints of general relativity \cite{Kodama90}. Remarkably, this Chern-Simons-Kodama (CSK) state yields a semiclassical solution that describes gravitons propagating on an inflating (de Sitter) background \cite{freidel2004linearization, Smolin02}.  It was shown in \cite{Alexander:2003wb} that this framework can be extended to inflation driven by a scalar field. Moreover, it has been established that  the CSK state reproduces both the Vilenkin and Hartle-Hawking wavefunctions of the universe under Fourier transform depending on the choice of integration contour for the Ashtekar variable \cite{magueijo2020equivalence, alexander2021generalized, magueijo2021real, paternoga2000triad}.  These cosmological solutions were extended to include fermions, leading to a resolution of the big-bang singularity\cite{Howard}.  These results promise a valuable connection between a non-perturbative statement of quantum gravity and the real universe contained as a semiclassical approximation. 
However, some serious problems linger with the CSK state: it is not invariant under large gauge transformations or CPT transformations---the latter of which precludes energy positivity---and perhaps worst of all, it is not normalizable with respect to the standard inner product \cite{Witten03, marugan1995exponential, freidel2004linearization}.

While the full Lorentzian CSK state requires a non-perturbative inner product, previous studies have demonstrated that the linearized CSK state is non-normalizable. The mechanism for the instability is simple: The perturbations split into positive and negative helicity states and while the positive helicity states have positive energy, the negative helicity ones  correspond to negative energy states \cite{freidel2004linearization}.  
One can also  speculate that the non-normalizibility of the CSK state is related to the issue of the quantum infrared-driven instability of de Sitter space\cite{Polyakov,Tsamis:1996qm,Anderson,Akh}. Subsequent studies argued that the negative frequency tensor perturbations about the CSK state correspond to anti-gravitons\cite{JB,BJ}.  Half of the full set of these modes corresponds to negative energy states.  The authors propose a physical condition that sub-horizon momentum modes that propagate perpendicular to their polarization should only exit, leading to the projecting out of negative energy modes\footnote{This physical condition translates into no graviton-mediated particle production for sub-horizon modes.  However, this condition of particle production will exist for super-horizon modes. }.  This condition is consistent with the CMB fluctuations considered in standard cosmological perturbation theory, and defines a positive energy, linearized physical inner product for gravitons.  Nonetheless, if these negative energy states do persist beyond quadratic order, it is imperative to identify a non-perturbative inner product, if one exists.

The results we present in this article do not fully resolve these issues. However, by identifying an alternative, non-perturbative inner product that follows naturally from reality conditions on the quantum Ashtekar connection, we pave the way for a non-perturbative analysis of these questions. In particular, the possibility to choose integration contours different from the naive ones may offer a possible resolution to the issue of non-normalizability of the CSK state. 
In section II, we provide a review of the CSK state from a covariant formulation. In section III, we define the new inner product and prove that it implies the reality conditions of the connection operator.  In section IV, we compute the determinant of the operator that determines the effective action.  We conclude with a discussion of open issues and future directions to be pursued. 

\section{Review of the CSK State}
Here we give a brief review of the CSK state, and show that it is annihilated by the quantum Hamiltonian constraint for a particular choice of factor ordering. Much of what appears in this section can also be found in \cite{Howard} and similar discussions appear in many other places, but we repeat it here for the purposes of making the article as self-contained as possible.\footnote{Note that the definition of the Planck length $\ell_{\mathrm{Pl}}$ used in this section and throughout the article differs from the one used in \cite{Howard} by a constant numerical factor, a choice that we make for the sake of notational convenience.} Our starting point in deriving the CSK state is the four-dimensional Holst Action, which is an extension of the Einstein-Cartan action of General Relativity, with a topological term: 
\be\label{act}
S=\int_{{\cal M}_4} [*(e^I\wedge e^J)\wedge R_{IJ}  + \rmi  e^I\wedge e^J\wedge R_{IJ}]\,,
\ee
where $*$ is the Hodge dual, the first term is the Einstein-Cartan action, and the second is the Holst term, which is, after integration by part, proportional to the first Bianchi identity.  Gravitational dynamics on a four-dimensional manifold ${\cal M}_4$ is described by the frame field $e_\mu^I(x)$, mapping a vector $v^\mu$ in the tangent space of ${\cal M}_4$ at the point $x$ into Minkowski spacetime $M_4$ (with metric $\eta_{IJ}={\rm diag}(-1,1,1,1)_{IJ}$). 
In this formulation, the Lorentz connection $\omega_{\mu I}{}^J$ is initially taken to be independent of the frame field. Its equation of motion imposes that it becomes torsionless.

Here we are interested in the Hamiltonian formulation in Ashtekar variables. In the Hamiltonian formulation one chooses a codimension one slice $\Sigma$ of spacetime and expresses the gravitational phase space variables as pull-backs on the hypersurface. One also imposes the normal gauge $e^0_a=0$, where $a,b,c$ denote indices tangent to $\Sigma$. This forces the coframe $e^0$ to be the hypersurface normal. The frame field variable is then encoded into their value on $\Sigma$. It is convenient to repack this information into  the densitized triad defined by $E^{a}_{i} = \frac{1}{2}\epsilon_{ijk} \epsilon^{abc} e^j_b e^k_c$. This quantity   is conjugate to the pull-back of the self-dual connection on $\Sigma$
\be
A_a^i(x)\equiv-\tfrac12 \e^{ij}_{\phantom{ij}k}\omega_{a j}^{\phantom{a j}k}-\rmi \omega_{a 0}^{\phantom{a 0}i}. \label{Ashtekar}
\ee
Since the Lorentz connection (and, in particular, the spin connection $\Gamma_\a^i\equiv -\tfrac12 \e^{ij}_{\phantom{ij}k}\omega_{\a j}^{\phantom{\a j}k}$) is real in real gravity, $A$ is complex-valued and obeys a non-trivial reality condition.
After imposition of the torsion constraints, the real part of $A$ is determined by the hypersurface triad, while the imaginary part is proportional to the extrinsic curvature of $\Sigma$ \cite{thiemann2008modern}. In other words we have the reality condition
\be\label{rc}
\bar{A}_a^{i}+ A_a^i=2\Gamma_a^i[E]\,,
\ee
where $\bar{A}_\a^{i}$ denotes the complex conjugate of ${A}_\a^{i}$, and $\Gamma[E]$ denotes the triad  spin connection which solves the equation $$\rmd e^i+\epsilon^i{}_{jk}\Gamma^j[E]\wedge e^k=0.$$
The Poisson bracket of the elementary variables $A$ and $E$ is
\be\label{comm}
\{A_a^i({\bf x}),\,E_j^b({\bf y})\}=\rmi G\delta_a^b\delta_j^i\delta({\bf x}- {\bf y})\,.
\ee
Introducing the ``magnetic'' field and the gauge field strength
\bea
B^{a i}&\equiv& \tfrac12\e^{abc}F_{bc}^i\,,\\
F_{ab}^k &=& \p_a A_b^k-\p_b A_a^k+\e_{ij}\phantom{}\phantom{}^{k}A_a^iA_b^j \,,\label{F}
\eea
one can show that the Hamiltonian scalar constraint following from Eq.~\Eq{act} is 
\bea
{\cal H}&\equiv& \e_{ijk} E^i\cdot E^j\times \left(B^k+\frac{\Lambda}3E^k\right)\nonumber\\
&=& \e_{ijk} E^{a i}E^{b j} \left(F_{ab}^k+\frac{\Lambda}{3}\e_{abc}E^{c k}\right)=0,\label{sca}
\eea
where $\Lambda$ is a cosmological constant (of any sign) and $\times$ is the vector spatial product defined as $({\bf a}\times {\bf b})_a= \e_{abc}a^b b^c$. In exterior algebra notation, the gauge field is $A\equiv A_\a \rmd x^\a\equiv A_\a^i \tau_i \rmd x^\a$ ($\tau_i$ being an $su(2)$ generator), the covariant derivative is $D=\rmd+A\wedge$, and Eq.~\Eq{F} can be compactly recast as $F=\rmd A+A\wedge A$. Under a local gauge transformation, the Ashtekar connection transforms as
\be\label{gt}
A \rightarrow A'= gAg^{-1} -g^{-1}\rmd g\,.
\ee
The full invariance group of the theory is the semidirect product of the diffeomorphism and gauge groups. Invariance under small gauge transformations is guaranteed by the Gauss constraint:
\be
{\cal G}_i
\equiv D_a E^a_i=\p_a E^a_i+ \e_{ijk}A_a^j E^{a k}=0\label{gau},
\ee
while spatial diffeomorphism invariance is imposed by the vector constraint
\be
{\cal V}_a \equiv (E_i\times B^i)_a =  F_{ab}^iE_i^b=0\,\label{vec}.
\ee
The total Hamiltonian is a linear combination of the constraints: up to constants, we have 
\be
H=G^{-1}\int_{\Sigma} \rmd^3x (N{\cal H}+\lambda^j{\cal G}_j+N^a{\cal V}_a),
\ee
where $\Sigma$ is the spatial submanifold and $N$, $\lambda^j$, and $N^\a$ are Lagrange multipliers.

Now let us construct the CSK state by solving the Wheeler-DeWitt equation.  We have the Hamiltonian (constraint)
\begin{align}
	{\cal H}_{WdW} = \epsilon_{ijk}E^{ai}E^{bj}(F^k_{ab} + \frac{\Lambda}{3}\epsilon_{abc}E^{ck}) = 0,
\end{align}
which acts on some wave function $\psi[A]$, and we want to find the form of $\psi[A]$ that satisfies the above relation. Applying the standard canonical quantization procedure, we get
\begin{align}
	\hat{E}^{ai}\rightarrow \ell^2_{\mathrm{Pl}}\frac{\delta}{\delta A_{ai}},
\end{align}
where $\ell^2_{\mathrm{Pl}}=\hbar G$.
The annihilation of the quantum state becomes
\begin{align}
	\widehat{\cal{H}}_{WdW}\,\psi[A] = \ell^4_{\mathrm{Pl}}\epsilon_{ijk}\frac{\delta}{\delta A_{ai}}\frac{\delta}{\delta A_{bj}}\Big(F^k_{ab} + \frac{\ell^2_{\mathrm{Pl}}\Lambda}{3}\epsilon_{abc}\frac{\delta}{\delta A_{ck}}\Big)\psi[A] = 0.
\end{align}
Of course, the Wheeler-DeWitt equation will be satisfied if the self-duality condition 
\be
\Big(F^k_{ab} + \frac{\ell^2_{\mathrm{Pl}}\Lambda}{3}\epsilon_{abc}\frac{\delta}{\delta A_{ck}}\Big)\psi[A] = 0
\ee
is satisfied, and the latter equation has a unique solution:
\be
\Psi_{CSK}[A] = \mathcal{N}e^{-\frac{3}{\ell^2_{\mathrm{Pl}}\Lambda} Y_{CS}[A]},
\ee
where $Y_{CS}[A]$ is the Chern-Simons functional
\be 
Y_{CS}[A] = \int \tr \left(\tfrac12 A\wedge dA + \tfrac{1}{3}A\wedge A\wedge A\right).
\ee 
\section{Inner product from reality conditions}

To summarize, we work, at the quantum level,  in the polarisation where the Ashtekar connection is diagonal. In this polarisation, the states are $\Psi(A)$ and the operators are given by 
 \be\label{opdef}
 \hat{E}^a_i \Psi = \ell^2_{\mathrm{Pl}} \frac{\delta \Psi }{\delta A_a^i}, \qquad 
 \hat{A}_a^i \Psi(A) = {A}_a^i \Psi(A).
 \ee
 We now  want to implement the reality conditions at the quantum level
 \bea\label{real}
 (\hat{E}^a_i)^\dagger&=&\hat{E}^a_i, \label{ReE}\\
 \hat{A}_{a}^i +(\hat{A}_{a}^i)^\dagger &=& 2\hat\Gamma_a^i. \label{ReA}
 \eea
 Here  $\hat{\Gamma}_a^i:=\Gamma_a^i(\hat{E})$ is the  operator corresponding to the 3d spin connection
 $\Gamma_a^i(E)$ which is the spin connection preserving the frame $e$: 
 \be
 \rd e^i + [\Gamma, e]^i=0. \label{FirstCartan}
 \ee
 The Lie algebra commutators are SL$(2,\mathbb{C})$ commutators given by $[A,B]^i:=\epsilon^{i}_{\,jk} A^j B^k.$
 
The main claim of this article is that the scalar product solving the reality condition is given by 
\be 
\langle  \Psi |\Psi \rangle   = 
 \int_{{\cal{C}}\times \bar{\cal{C}}}\!\!\! DA D \bar{A} \,\, \exp\left({- S(\Re A) } \right) \left|\Psi(A)\right|^2,
\ee
where we denote by $\Re A = \frac12 ( A+\bar{A})$, the real part of the self-dual connection $A$.
We assume that the integration contours $\cal C$ and $\bar{\cal C}$ for the holomorphic and antiholomorphic components allow integration by parts: \textit{i.e.}, they are such that 
\be
\int_{\cal C} DA \frac{\delta \Psi(A)}{\delta A_a^i} =0 = \int_{\bar{\cal C}} D\bar{A} \frac{\delta \bar{\Psi}(\bar{A})}{\delta \bar{A}_a^i}\,.
\ee 
It is important to note that these are the only conditions necessary for our proof of the reality condition, leaving us some freedom in the choice of contours that could be used to fix the normalizability issue.
Finally, $S(\Re A)$ is an effective action defined by the following path integral
\be
\exp\left(-  S(\Re A) \right) := \int DE\, \exp\left( - \frac1{\ell_{\mathrm{Pl}}^2} \int e_i \wedge \rd_{\Re A} e^i \right), \label{effAc}
\ee
where $ \rd_A e^i = \rd e^i + [A,e]^i$.
The main argument is that the first reality condition \eqref{ReE} forces the measure to depend only on $\Re A$,
while the second reality condition \eqref{ReA}  uniquely determines the form of the effective action $S(\Re A)$.

The detailed proof of the statement goes as follows:
first one uses that 
\bea
\frac{\delta}{\delta A_a^i} \left(  \int_\Sigma e_i \wedge \rd_{\Re A} e^i \right)& =&    E_i^a= \frac{\delta}{\delta \bar{A}_a^i} \left(  \int_\Sigma e_i \wedge \rd_{\Re A} e^i \right),\label{varA}\\
\frac{\delta}{\delta E_i^a} \left(  \int_\Sigma e_i \wedge \rd_{\Re A} e^i \right)& =&   A_a^i + \bar{A}_a^i - 2 \Gamma_a^i(E).\label{varE}
\eea
A proof of the second equality is given in appendix \ref{varapp}.
From this and the definition \eqref{opdef} of the operators $(\hat{A}_a^i,\hat{E}^a_i)$, we get that 
\be\label{herm1}
\left(\frac{\delta }{\delta \bar{A}_a^i} - \frac{\delta }{\delta A_a^i} \right) e^{- S(\Re A)} = 0,
\ee
and
\bea
\label{herm2-1}
\left[A_a^i +\bar{A}_a^i -2 \hat{\Gamma}_a^i(-\hat{E})\right] e^{- S(\Re A)} &=&
\int DE ( (A_a^i +\bar{A}_a^i -2 \hat{\Gamma}_a^i(E)) \exp\left(-\frac1{\ell_{\mathrm{Pl}^2}}\int_\Sigma e_i \wedge \rd_{\Re A} e^i \right)\cr 
&=&
\int DE \frac{\delta }{\delta {E}^a_i} \exp \left(  -\frac1{\ell_{\mathrm{Pl}^2}}\int_\Sigma e_i \wedge \rd_{\Re A} e^i \right) =0. \label{herm2-2}
\eea
From which it follows simply that 
\bea
\langle  \Psi|  \hat{E}^a_i- (\hat{E}^a_i)^\dagger|\Psi \rangle&=&\ell^2_{\mathrm{Pl}}
 \int_{{\cal{C}}\times \bar{\cal{C}}} DA D \bar{A} \,  e^{-S(\Re A)} \left(\Psi(\bar{A}) \frac{\delta \Psi (A)}{\delta A_a^i}-\frac{\delta \Psi (\bar{A})}{\delta \bar{A}_a^i}\Psi(A)\right) \cr
 &=& \ell^2_{\mathrm{Pl}} \int_{{\cal{C}}\times \bar{\cal{C}}} DA D \bar{A} \, \left(\frac{\delta e^{-S(\Re A)}}{\delta \bar{A}_a^i} - \frac{\delta e^{-S(\Re A)}}{\delta A_a^i} \right)
 |\Psi(A)|^2 =0,
\eea
and that 
\bea
\langle  \Psi|  \hat{A}_a^i +(\hat{A}_a^i)^\dagger|\Psi \rangle&=&
 \int_{{\cal{C}}\times \bar{\cal{C}}} DA D \bar{A} \, (A_a^i +\bar{A}_a^i ) e^{-S(\Re A)} |\Psi(A)|^2 \cr
 &=& \int_{{\cal{C}}\times \bar{\cal{C}}} DA D \bar{A} \, \left(2 {\Gamma}_a^i(-\hat{E}) e^{-S(\Re A)} \right) 
 |\Psi(A)|^2.
\eea
In the last line, the operator $\hat{E}_i^a$ can be represented as either $\ell^2_{\mathrm{Pl}} \frac{\delta }{\delta A_a^i}$ or $\ell^2_{\mathrm{Pl}}\frac{\delta }{\delta \bar{A}_a^i}$ since we have  the Hermiticity condition \eqref{herm1}. 
Integrating by parts, we obtain that this is equal to 
\be
\int_{{\cal{C}}\times \bar{\cal{C}}} DA D \bar{A} \,  e^{-S(\Re A)}  
 \Psi^*(\bar{A}) \left(2 {\Gamma}_a^i(\hat{E}) \Psi(A)\right).
\ee
This concludes the proof. We see that the  inner product involves an effective action $S(\Gamma)$, which is evaluated as the log of a determinant in the next section. It also involves a choice of closed contours $(\cal{C}, \bar{\cal{C}})$ which  affects the definition of the product. 

\section{Evaluation of the effective action}
In this section, we seek to understand the properties of the effective action $S(\Gamma)$, where $\Gamma:=\Re A$  denotes the real part of the self-dual connection.
In order to do so, it is convenient to restrict our analysis to connections $\Gamma$
which are \textit{non-degenerate}. These satisfy the non-degeneracy condition
\be \label{nondeg}
\nu_{abc} B^a_i(\Gamma)B^b_j(\Gamma)B^c_k(\Gamma) \epsilon^{ijk} \neq 0, 
\ee
where $\nu_{abc}=\nu \epsilon_{abc}$, with $\nu>0$, is a volume form and $\nu_{abc} B^c_i(\Gamma) := R_{ab}^j(\Gamma)\delta_{ij}$ is the magnetic field associated with the curvature of $\Gamma$.

In appendix \ref{spinc} we show that when the connection is non-degenerate, there exists a frame field $e_\Gamma$ such that \be 
F(\Gamma) =\pm \frac{1}{2}[e_\Gamma,e_\Gamma],
\ee 
where $\pm$ is the sign of \eqref{nondeg}.
The frame field $e_\Gamma$ is uniquely defined by this condition.

Since the action $\int (e_i\wedge \rd_\Gamma e^i)$ is purely quadratic, the effective action simply computes the determinant of $\rd_\Gamma$. More precisely, the operator $\rd_\Gamma: \Omega_{1}(\mathfrak{su}(2)) \to \Omega_{2}(\mathfrak{su}(2))$ maps Lie algebra valued one-forms onto Lie algebra valued two forms. We can compose this operation with the Hodge duality operator $*_\Gamma:\Omega_{2}(\mathfrak{su}(2)) \to \Omega_{1}(\mathfrak{su}(2))$ which is constructed from the metric $(q_\Gamma)_{ab} = (e_\Gamma)_a^i(e_\Gamma)_b^j\delta_{ij}$.
This gives an endomorphism  $D_\Gamma : \Omega_{1}(\mathfrak{su}(2)) \to \Omega_{1}(\mathfrak{su}(2))$ where $D_\Gamma =*_\Gamma \rd_\Gamma$, which is explicitely given by
    \be 
    D_\Gamma : e_a^i \to (D_\Gamma e)_a^i 
    = \nu_a{}^{bc}\nabla^\Gamma_b e_c^i\,.
    \ee 
Here $\nu_\Gamma=\nu_{abc} \,\rd x^a\wedge \rd x^b \wedge \rd x^c$ is the measure associated with $e_\Gamma$, while spacetime indices are raised with the metric $q_\Gamma$.
Given this definition, we then get that 
\be
e^{ S(\Gamma)} =  \frac{\mathrm{det}( D_\Gamma)}{\mathrm{det}(e_\Gamma)}.
\ee 

The goal is now to evaluate this determinant in a homogeneously curved background where $(\Gamma_R)_a^i := R \delta_a^i$ is constant. The simplest background that we can consider is the flat slicing of de Sitter. In this case, the determinant can be computed by decomposing the fields in Fourier modes. However, in a flat background $D_\Gamma$ is completely singular, so we consider a closed slicing of de Sitter with spacetime metric
\be
ds^2 = -dt^2 + a(t)^2d\Omega_3^2
\ee
where $a(t)^2 = \frac{3}{\Lambda}\cosh^2\left(\sqrt{\frac{\Lambda}{3}}t\right)$, and $d\Omega_3^2$ is the round metric on the three-sphere $\mathbb{S}^3$. A convenient orthonormal frame for the unit three-sphere is given by the left-invariant one-forms of SU(2)
\begin{eqnarray*}
e_0^1 &=& \frac{1}{2}(\cos\psi d\theta + \sin\psi\sin\theta d\phi) \\
e_0^2 &=& \frac{1}{2}(\sin\psi d\theta - \cos\psi\sin\theta d\phi) \\
e_0^3 &=& \frac{1}{2}(d\psi + \cos\theta d\phi).
\end{eqnarray*}
This frame satisfies $de_0^i = \epsilon^i_{\,jk} e_0^j \wedge e_0^k.$  We can define a physical frame $e^i = a(t)e_0^i$, which satisfies $de^i = \frac{1}{a}\epsilon^i_{\,jk}e^j\wedge e^k$, from which we can read off the spin connection $\omega^i_{\, j} = \frac{1}{a}\epsilon^i_{\, jk}e^k$. Using $\Gamma^i = -\frac{1}{2}\epsilon^{ijk}\omega_{jk}$ leads to 
$$\Gamma^i = -\frac{1}{a}e^i.$$ In this coordinate system, the extrinsic curvature is just the time derivative of the spatial metric, so we have $K_{ij} = 2H\delta_{ij}$, with $H = \frac{\dot a}{a}$, which leads to
$$K^i = 2He^i.$$ 
So the Ashtekar connection for de Sitter in closed slicing is 
$$A^i = \left(-\frac{1}{a} + 2iH\right)e^i,$$
or
$$A^i = \left(-1 + 2i\dot{a} \right)e_0^i.$$
It can be verified that the associated spin connection satisfies the non-degenaracy condition \eqref{nondeg}, so its existence is an important check on the self-consistency of our approach.

Now, we would like to compute the determinant of the operator $D_\Gamma$ defined by
\be
D_\Gamma X^i{}_a = \epsilon_a{}^{bc}\nabla_b X^i{}_c.
\ee
Since any Lie algebra-valued one-form can be identified with a covariant, rank two tensor via ${X_{ab} = e_{ia}X^i{}_b}$, we can equivalently view $D_\Gamma$ as an endomorphism on rank two tensors:
\be
D_\Gamma X_{ab} = \epsilon_a{}^{cd}\nabla_c X_{db}.
\ee

Note that this operator has a kernel since $D_\Gamma g_{ab} = 0$. We therefore expect that its determinant is zero, and we verify this explicitly below. However, we can still calculate a non-zero pseudo-determinant for this operator by considering the product of its non-zero eigenvalues, and this is what we would expect the path integral to compute. The details of this procedure can also be found in appendix \ref{BigCalc}, where we diagonalize $D_\Gamma$ by expanding it in vector and tensor harmonics on $\mathbb{S}^3$. The three-sphere spherical Harmonics  $Y_{k\ell m}$ depend on 3 integers  $k,\ell$, and $m$, with $k \geq \ell \geq 0$ and  $\ell\geq m\geq -\ell$. In coordinates where the three-sphere metric is $\rd s^2 =\rd \chi^2 +\sin^2\chi(\rd\theta^2 +\sin^2\theta \rd\varphi^2)$, we have that  $(\ell, m)$ are the usual two-sphere spherical harmonics that controls the dependence on $(\theta,\varphi)$, while $k$ is the quantum number that determines the value of the Laplacian and the dependence on the angle $\chi$. On the unit sphere we have $\nabla_a\nabla^a Y_{k\ell m}=-k(k+2) Y_{k\ell m}$. 

To diagonalize $D_\Gamma$ we have to 
expand the 2-tensors $X_{ab}$ in terms of 2-tensor Harmonics. Since there are 9 such harmonics, we have nine types of expansion coefficients $\tilde{X}_{(A)}^{\klm}$, $A = 0,\ldots,8$ labelled by the spherical harmonics numbers $(k,\ell,m)$.
One finds that like the Laplacian, the operator $\tilde{D}_\Gamma$---i.e., the coefficients of the operator $D_\Gamma$ written in the basis of tensor harmonics on the round three-sphere---is simply a $9\times 9$ matrix which depends on $k$, but not $\ell$ or $m$.   The main result of appendix \ref{BigCalc} is the following system of linear equations:
\begin{eqnarray}
\tilde{D}_\Gamma\tilde{X}^{k\ell m}_{(0)}
&=& -\frac{\sqrt{k(k+2)}}{\sqrt{3}a}\tilde{X}^{k\ell m}_{(6)}, \nonumber \\
\tilde{D}_\Gamma\tilde{X}^{k\ell m}_{(1)} &=& \frac{k+1}{2a}\tilde{X}^{k\ell m}_{(2)} +\frac{2 - (k+1)^2}{a\sqrt{2(k-1)(k+3)}}\tilde{X}^{k\ell m}_{(7)}, \nonumber \\
\tilde{D}_\Gamma\tilde{X}^{k\ell m}_{(2)} &=& \frac{k+1}{2a}\tilde{X}^{k\ell m}_{(1)} + \frac{2 - (k+1)^2}{a\sqrt{2(k-1)(k+3)}}\tilde{X}^{k\ell m }_{(8)} , \nonumber \\
\tilde{D}_\Gamma\tilde{X}^{k\ell m}_{(3)} &=& \frac{\sqrt{(k-1)(k+3)}}{\sqrt{6}a}\tilde{X}^\klm_{(6)}, \nn \\
\tilde{D}_\Gamma\tilde{X}^{k\ell m}_{(4)} &=& \frac{k+1}{a}\tilde{X}^{\klm}_{(5)}, \nn \\
\tilde{D}_\Gamma\tilde{X}^{k\ell m}_{(5)} &=& \frac{3k(k+2)}{2a(k+1)}\tilde{X}^\klm_{(4)}, \nn \\
\tilde{D}_\Gamma\tilde{X}^\klm_{(6)} &=& -\frac{2\sqrt{k(k+2)}}{\sqrt{3}a}\tilde{X}^\klm_{(0)} - \frac{\sqrt{2(k-1)(k+3)}}{\sqrt{3}a}\tilde{X}^\klm_{(3)}, \nn \\
\tilde{D}_\Gamma\tilde{X}^\klm_{(7)} &=& -\frac{\sqrt{(k-1)(k+3)}}{\sqrt{2}a}\tilde{X}^\klm_{(1)} + \frac{1 + k(k+2)}{a(k+1)}\tilde{X}^\klm_{(8)}, \nn \\
\tilde{D}_\Gamma\tilde{X}^\klm_{(8)} &=& -\frac{\sqrt{(k-1)(k+3)}}{\sqrt{2}a}\tilde{X}^\klm_{(2)} + \frac{1 + k(k+2)}{a(k+1)}\tilde{X}^\klm_{(7)},
\end{eqnarray}
which defines the matrix $\tilde{D}_\Gamma$.
As promised, $\tilde{D}_\Gamma$ is singular---one of its eigenvalues is zero, while the remaining eight are non-zero. In particular, the eigenvalues are $\{0, \pm \lambda_1, \pm\lambda_2, \pm \lambda_{\pm}\}$, where
\bea
\lambda_1 &=& {\frac {1}{a}\sqrt {{k}^{2}+2\,k-1}}, \nn \\
\lambda_2 &=& {\frac {1}{2\,a}\sqrt {6\,{k}^{2}+12\,k}}, \nn \\
\lambda_\pm &=& {\frac {1}{4\,a}\sqrt {18\,{k}^{2} + 36\,k + 2 \pm 6\,\sqrt {9
\,{k}^{4}+36\,{k}^{3}+38\,{k}^{2}+4\,k-7}}}\,.
\eea 
The product $\tilde{\Delta}_k$ of the eight non-zero eigenvalues has a simple form:
\be 
\tilde{\Delta}_k = \frac{3}{2a^8}\left[(k+1)^2 - 1\right]\left[(k+1)^2 - 2\right].
\ee 
The na\"{i}ve pseudo-determinant of $D_\Gamma$ is then a product over $k, \ell, m$ of this quantity:
\bea 
\prod_\klm \tilde{\Delta}_k 
= \prod_{k = 0}^\infty\tilde{\Delta}_k^{(k+1)^2},
\eea
where we used that the dimension of the $k$ representation is $\sum_{\ell=0}^k(2\ell+1)=(k+1)^2$.
However, this quantity also vanishes, since $\tilde{\Delta}_0 = 0$. We therefore truncate the product to positive values of $k$ to define the pseudo-determinant of $D_\Gamma$:
\be
\psi\!\det(D_\Gamma) = \prod_{k = 1}^\infty\tilde{\Delta}_k^{(k+1)^2}.
\ee 
Finally, we can re-index the product by defining $n = k+1$ to obtain
\be
\psi\!\det(D_\Gamma) = \prod_{n = 2}^\infty\left[\frac{3}{2a^8}(n^2 - 1)(n^2 - 2)\right]^{n^2}. \label{FinalAnswer}
\ee
This expression needs to be regulated, as we now discuss.
\section{Discussion}
We have shown that a non-trivial inner product measure for the states of four-dimensional quantum gravity follows from a simple reality condition on the quantum Ashetkar connection, and a closure assumption about its integration contour,  leaving room for the possibility that the CSK state might be normalizable when the correct inner product associated with the right choice of the contour is taken into account. While we have taken some first steps toward evaluating the inner product on a homogeneously curved slice of de Sitter space, which is an appropriate setting for studying the CSK state which has de Sitter as a semi-classical limit, the work we have presented here is far from conclusive. Indeed, even the simplified problem of studying the inner product on a homogeneous background proved to be a non-trivial task. While we have attempted to keep the presentation of the main results as clean as possible, the careful reader will appreciate that the calculation of the expansion coefficients in appendix \ref{BigCalc} was a laborious undertaking, and even then the resulting infinite product is divergent.

Nevertheless, it may be possible to find a finite determinant without specifying a background by performing a zeta-function regularization for the determinant of $D_\Gamma$. This prescription was first used by Hawking to regulate the path integral of Euclidean gravity on de Sitter space \cite{Hawking:1976ja}.  Given that the determinant is in the form of a divergent product of eigenvalues $\det(D_\Gamma) = \prod_n \lambda_n $, one can form a zeta function from the eigenvalues $\lambda_n(D_\Gamma) $ of the operator $D_\Gamma$:
\be 
\zeta(s,D_\Gamma) = \sum_{n}\lambda^{-s}_{n}(D_\Gamma).
\ee 
One would expect such an expansion to converge for $\Re(s) > 2$, and to define via analytic continuation a meromorphic function of $s$ on the whole complex plane that is analytic at $s=0$.  We will pursue this procedure in an upcoming work.

Another fascinating possibility is provided by Witten's fundamental study of Chern-Simons theory for complex groups \cite{Witten:2010cx}. From his work, we know that the proper definition of the Chern-Simons partition function  requires a lucid understanding of what integration contour is used in defining the theory. From our analysis, it is clear that there is a connection between computing the norm of the CSK state in quantum gravity and computing the Chern-Simons partition function for a complex group. We plan to study this connection  in a forthcoming work. The issue of contour integration for the CSK state was already touched on in \cite{magueijo2020equivalence} in a mini-superspace context.  

As we have noted, and as is often the case with operator determinants derived from path integrals in quantum field theories, the product \eqref{FinalAnswer} diverges. In order to go further, one needs a prescription for either regulating the operator determinant (such as a zeta-function regularization), or interpreting it as the divergent part of some physically meaningful quantity. In the present case, the latter possibility seems promising, since we have only evaluated $\psi det(D_\Gamma)$ on a slice of de Sitter space, while the full, non-perturbative inner product measure requires us to path-integrate over all possible configurations. It could well be the case that the pseudo-determinant we have computed for the round three-sphere might serve as a regulator for a perturbative treatment of the inner product measure expanded about this homogeneously curved background in the same manner that the path integral for a quantum field theory without sources serves as a regulator for the generating functional including sources. However, even a perturbative treatment of the inner product measure about the round three-sphere presents significant challenges, since one either needs to develop a perturbation theory for the vector and tensor harmonics, or else come up with another approach altogether for computing the measure perturbatively.

Yet another promising direction that could allow us to make progress with computing the inner product measure is rooted in the Belinskii Khalatnikov Lifshitz (BKL) conjecture, which states that as one approaches a space-like singularity, spatial derivatives become negligible \cite{belinskii1970oscillatory}.  In the context of the Kodama state, this means that only the cubic, non-abelian part of the action will be non-vanishing.  This was already confirmed in previous work where the Kodama state was evaluated in Bianchi-type spacetimes and it was shown that only the non-Abelian term in the action functional is present\cite{alexander2021generalized}.  Conversely, it was also shown that when the spacetime is homogeneous and isotropic, the non-Abelian term in the action vanishes.  This is reminiscent of non-abelian projection proposed by t' Hooft, to exploit features of confinement of gauge theories \cite{tHooft:1981bkw}. We speculate that in the BKL limit, as the spacetime enters a space-like singularity, this mix-master behavior, proposed to avoid singularities, could be described by a Chern-Simons matrix theory, which could perhaps be obtained from our proposed inner product.

Many of the ingredients in the approach we have taken in this article also appear in a series of papers by Haggard, Han, Kami\'{n}ski and Riello on $SL(2,\mathbb{C})$ Chern-Simons theory and four-dimensional quantum gravity with a cosmological constant \cite{haggard2015sl-1, haggard2015sl-2, haggard2016four}, and it would be interesting to develop those connections further. Finally, any observer in a universe with a positive cosmological constant is limited by their cosmological horizon. It would be interesting to consider how the inner product we have introduced here might affect the analysis of local sub-systems as in \cite{donnelly2016local} and entanglement entropy \cite{donnelly2008entanglement} for the Kodama state.

\section{Acknowledgments}
We are grateful to Jim Gates, Ibou Bah, Jo\~{a}o Magueijo, Jim Simons, Dan Freed, Edward Frenkel, Steve Carlip, and Lee Smolin for discussions and helpful feedback. 

Research at Perimeter Institute is partly supported by the Government of Canada through the Department of Innovation, Science and Economic Development Canada and the Province of Ontario through the Ministry of Colleges and Universities.
\appendix

 \section{Evaluation of variations }\label{varapp}
 In this section we give the proof of the variational derivative (\ref{varA},\ref{varE})

 Using that $ E^i= \frac12 [e,e]^i$, we have.
\bea
\delta\left(  \int_\Sigma e_i \wedge \rd_a e^i \right)&=&  
2 \int_\Sigma \delta e_i \wedge \rd_a e^i + \int_\Sigma  e_i \wedge [\delta a, e]^i\cr
&=& 
2\int_\Sigma \delta e_i \wedge [a-\Gamma,  e]^i + \int_\Sigma \delta a_i \wedge [e, e]^i \cr
&=&2 \int_\Sigma [\delta e, e]_i \wedge (a-\Gamma)^i + 2 \int_\Sigma \delta a_i \wedge E^i\cr
&=& \int_\Sigma \delta E_i \wedge (2a-2\Gamma)^i + 2\int_\Sigma \delta a_i \wedge E^i.
\eea
we used that 
\be 
\delta E^i =\delta \left(\frac12 [e,e]^i\right) = [\delta e, e]^i.
\ee 

\section{Spin connection}\label{spinc}
In this section we show that given a connection $\Gamma$ we can construct a frame field $e_\Gamma$ solution of $\rd_\Gamma e_\Gamma=0$ \cite{Herfray:2016std}. 
Given a 3d real connection $\Gamma_a^i$ we define its curvature by $F(\Gamma)= \rd \Gamma+ \epsilon^i{}_{jk} \Gamma^j\wedge \Gamma^k$.
Given a volume form $\bm\nu= \nu_{abc} \rd x^a\wedge \rd x^b\wedge \rd x^c =\nu \rd^3 x $ we define
the tensor $B^a_j$ to be such that
\be
F_{ab}^i = \nu_{abc} B^c_j \delta^{ij}. 
\ee 
From this we define
\be
{\det}_\nu(B):=\frac16 \nu_{abc} B^a_iB^b_jB^c_k \epsilon^{ijk}.
\ee
under rescaling of the volume form $\bm\nu\to\lambda \bm\nu$ we have 
\be
  B^a_i \to \lambda^{-1}  B^a_i, \qquad 
  {\det}_\nu( B)\to \lambda^{-2}{\det}_\nu( B).
\ee
this means that $s_\Gamma=\mathrm{sign}(\det( B))$ is independent of the choice of background volume form and we  define a volume form $\nu_\Gamma$ to be such that 
\be
\nu_\Gamma := \sqrt{|{\det}_\nu( B)|} \nu.
\ee
$\nu_\Gamma$ is independent of the background volume form. 

We can now define a frame field $e_\Gamma$ such that 
\be
F(\Gamma):= \frac{s_\Gamma}2 [e_\Gamma,e_\Gamma] . 
\ee
In coordinates this means that 
\be 
\hat{e}^a_i = \frac{\nu}{\nu_\Gamma} B^a_i,\qquad 
\det(e) = s_\Gamma \nu_\Gamma.
\ee
where $\hat{e}_i$ is the frame field inverse to the form $e^i$.
This also means that
\be
 [F_{[ab}(\Gamma), F_{c ] d}(\Gamma)]^k:= S_\Gamma (\bm{\nu}_\Gamma)_{abc} ({e}_\Gamma)_d^k
\ee
\be
 \frac13 \frac{\nu}{\nu_\Gamma}  [B^c(\Gamma), F_{c d}(\Gamma) ]^k:=   (\hat{e}_\Gamma)_d^k
\ee
\section{Calculation of the pseudo-determinant of $D_\Gamma$ on $\mathbb{S}^3$} \label{BigCalc}
In order to diagonalize $D_\Gamma$, it's useful to introduce  the vector and tensor harmonicson $\mathbb{S}^3$, which, following the notations and conventions in \cite{lindblom2017scalar}, can be defined in terms of first derivatives of the scalar
harmonics $Y^{k\ell m
}$
as follows:
\begin{eqnarray}
Y^{k\ell m}_{(0)\,a}&=& \frac{a}{\sqrt{k(k+2)}}\nabla_aY^{k\ell m},
\label{e:Yklm(0)i} \\
Y^{k\ell m}_{(1)\,a}&=& \frac{a^2}{\sqrt{\ell(\ell+1)}}
\epsilon_{a}{}^{bc}\nabla_bY^{k\ell m}\nabla_c\cos\chi,
\label{e:Yklm(1)i} \\
Y^{k\ell m}_{(2)\,a}&=& \frac{a}{k+1}\epsilon_a{}^{bc}\nabla_bY^{k\ell m}_{(1)\,c},
\label{e:Yklm(2)i}
\end{eqnarray}
where $\epsilon_{abc}$ is the tensor volume
element associated with $g_{ab}$, and the definition of the angle $\chi$ appearing in the definition of $Y^{k\ell m}_{(1)\,a}$ can be found in \cite{lindblom2017scalar}, but is not important for our purposes. The vector harmonics $Y^{k\ell m}_{(A)\,a}$ are eigenfunctions of the
Laplacian on $\mathbb{S}^3$ with the following eigenvalues:
\begin{eqnarray}
  \nabla^b\nabla_bY^{k\ell m}_{(0)\,a}&=& \frac{2-k(k+2)}{a^2}Y^{k\ell m}_{(0)\,a},
  \label{e:EigVecY0}\\
  \nabla^b\nabla_bY^{k\ell m}_{(1)\,a}&=& \frac{1-k(k+2)}{a^2}Y^{k\ell m}_{(1)\,a},
  \label{e:EigVecY1}\\
  \nabla^b\nabla_bY^{k\ell m}_{(2)\,a}&=& \frac{1-k(k+2)}{a^2}Y^{k\ell m}_{(2)\,a}.
  \label{e:EigVecY2}
\end{eqnarray}
For future convenience, we also record the divergence of each of the vector harmonics:
\begin{eqnarray}
  \nabla^aY^{k\ell m}_{(0)\,a}&=& -\frac{\sqrt{k(k+2)}}{a}Y^{k\ell m},
  \label{e:DivVecY0}\\
\nabla^aY^{k\ell m}_{(1)\,a}&=& 0\label{e:DivVecY1},\\
\nabla^aY^{k\ell m}_{(2)\,a}&=& 0\label{e:DivVecY2}.
\end{eqnarray}
Now we can express any one-form $V_a$ as an expansion in vector
harmonics:
\begin{eqnarray}
  V_a=\sum_{A k \ell m}
\tilde{V}^{k\ell m}_{(A)}
\,Y^{k\ell m}_{(A)\,a},\qquad
\end{eqnarray}
where the range of allowed values for $k$ depends on $A$, the range of allowed values for $\ell$ depends on $k$, and the range of allowed values for $m$ depends on $\ell$, but the details are not important at this point.
Using the orthonormality relation
\begin{eqnarray}
\delta_{(A)(B)}\delta^{\,kk'}\delta^{\,\ell\ell'}\delta^{\,mm'}=
\frac{1}{a^3}\int\!\! 
g^{ab}\,Y^{\,k\ell m}_{(A)\,a}\,Y^{\,*k'\ell' m'}_{(B)\,b}\!\sqrt{g}\,
d^{\,3}x,\!\!\!\!\!\nonumber
\label{e:VectorOrthoNormal}
\end{eqnarray}
the expansion coefficients can be written as
\begin{eqnarray}
\tilde{V}^{k\ell m}_{(A)}=\frac{1}{a^3}\int V^a \,Y^{*k\ell m}_{(A)\,a} \sqrt{g}\,d^{\,3}x.
\label{e:VectorExpansionCoefficients}
\end{eqnarray}
For future reference, we compute $\epsilon^{abc}\nabla_a Y^{k\ell m}_{(A)b}$ for $A = 0,1,2$. For $A = 0$, we have
\be
\epsilon^{abc}\nabla_a Y^{k\ell m}_{(0)b} = \epsilon^{abc}\nabla_a\Big(\frac{a}{\sqrt{k(k+2)}}\nabla_b Y^{k \ell m}\Big) = 0.
\ee
For $A = 1$, we have
\be
\epsilon^{abc}\nabla_a Y^{k\ell m}_{(1)b} = \frac{k + 1}{a}g^{bc}\,Y^{k\ell m}_{(2)b},
\ee
where we used equation \eqref{e:Yklm(2)i}.
For $A = 2$ we have to do a calculation:
\bea
\epsilon^{abc}\nabla_a Y^{k\ell m}_{(2)b} &=& \frac{a}{k+1}\epsilon^{bca}\epsilon_b{}^{de}\nabla_a\nabla_d Y^{k \ell m}_{(1)e} \nn \\
&=& \frac{a}{k+1}\left(\nabla^e\nabla^c Y^{k\ell m}_{(1)e} - g^{ce}\nabla^a\nabla_a Y^{k\ell m}_{(1)e}\right) \nn \\
&=& \frac{a}{k+1}\left[ R^{ec}{}_e{}^f Y^{k\ell m}_{(1)f} + \nabla^c\nabla^eY^{k\ell m}_{(1)e} - g^{ce}\left(\frac{1 - k(k+2)}{a^2}\right)Y^{k\ell m}_{(1)e} \right],
\eea
where in going from the second line to the third line we used equation \eqref{e:EigVecY1}. Now, the first term in the last line above is just the Ricci tensor, which for the three-sphere is proportional to the metric: $R^{ec}{}_e{}^f = R^{cf} = \frac{2}{a^2}g^{cf}$. Meanwhile, the second term in the last line vanishes due to equation \eqref{e:DivVecY1}. So we are left with 
\be
\epsilon^{abc}\nabla_a Y^{k\ell m}_{(2)b} = \frac{1 + k(k+2)}{a(k+1)}g^{ca}Y^{k\ell m}_{(1)a} = \frac{k+1}{a}g^{ca}Y^{k\ell m}_{(1)a}.
\ee
To summarize, we have
\bea
\epsilon^{abc}\nabla_a Y^{k\ell m}_{(0)b} &=& 0, \label{curl0} \\
\epsilon^{abc}\nabla_a Y^{k\ell m}_{(1)b} &=& \frac{k+1}{a}g^{ca}Y^{k\ell m}_{(2)a}, \label{curl1} \\
\epsilon^{abc}\nabla_a Y^{k\ell m}_{(2)b} &=& \frac{k+1}{a}g^{ca}Y^{k\ell m}_{(1)a}. \label{curl2}
\eea
The vector harmonics $Y^{k\ell m}_{(A)a}$ can also be used to expand two-forms $\alpha_{ab} = -\alpha_{ba}$:
\begin{eqnarray}
\alpha_{ab}=
  \sum_{A k \ell m}
\!\!\tilde{\alpha}^{k\ell m}_{(A)}\epsilon_{ab}{}^c Y^{k\ell m}_{(A)\,c}.\qquad
\end{eqnarray}
In order to expand general rank-two tensors, we need to introduce the (symmetric) tensor harmonics:
\begin{eqnarray}
&&
Y^{k\ell m}_{(0)\,ab}= \frac{1}{\sqrt{3}}Y^{k\ell m}g_{ab},
\label{e:Yklm(0)ij}\\
&&
Y^{k\ell m}_{(1)\,ab}= \frac{a}{\sqrt{2(k-1)(k+3)}}
\left(\nabla_aY^{k\ell m}_{(1)\,b}
+\nabla_bY^{k\ell m}_{(1)\,a}\right),
\label{e:Yklm(1)ij}\\
&&
Y^{k\ell m}_{(2)\,ab}= \frac{a}{\sqrt{2(k-1)(k+3)}}
\left(\nabla_aY^{k\ell m}_{(2)\,b}
+\nabla_b Y^{k\ell m}_{(2)\,a}\right),
\label{e:Yklm(2)ij}\\
&&
Y^{k\ell m}_{(3)\,ab}= \frac{\sqrt{3}a}{\sqrt{2(k-1)(k+3)}}
\Biggl(\nabla_a Y^{k\ell m}_{(0)\,b}
+\frac{\sqrt{k(k+2)}}{\sqrt{3}a}Y^{k\ell m}_{(0)\,ab}\Biggr),\\
\label{e:Yklm(3)ij}
&&
Y^{k\ell m}_{(4)ab}= a\sqrt{\frac{(\ell-1)(\ell+2)}{2k(k+2)}}\biggl\{\tfrac{1}{2} 
E^{k\ell}\left( \nabla_a F^{\ell m}_{b}
+\nabla_b F^{\ell m}_{a}\right)\nonumber\\
&& \qquad+ 
\csc^2\!\chi\left[\tfrac{1}{2}(\ell-1)\cos\chi\, E^{k\ell}+C^{k\ell}
\right]
\left(F^{\ell m}_{a}\nabla_b\cos\chi+ F^{\ell m}_{b}\nabla_a\cos\chi\right)
\biggr\},\quad
\label{e:Yklm(4)ij}\\
&&
Y^{k\ell m}_{(5)\,ab}= \frac{a}{2(k+1)}
\left(\epsilon_a{}^{ce}\nabla_c Y^{k\ell m}_{(4)\,eb} 
+\epsilon_b{}^{ce}\nabla_c Y^{k\ell m}_{(4)\,ea}\right),\qquad
\label{e:Yklm(5)ij}
\end{eqnarray}
where the quantities $E^{k\ell}, C^{k\ell}$, and $F_a^{\ell m}$ appearing in the definition of $Y^{k\ell m}_{(4)\,ab}$ are defined in \cite{lindblom2017scalar} (along with the angle $\chi$) but are not important for our purposes. \\ 
\indent The symmetric tensor harmonics $Y^{k\ell m}_{(A)\,ab}$ are eigenfunctions of the Laplacian with the following eigenvalues:
\bea
  \nabla^n\nabla_nY^{k\ell m}_{(0)\,ab}&=& -\frac{k(k+2)}{a^2}Y^{k\ell m}_{(0)\,ab},
  \label{e:EigTensorY0}\\
  \nabla^n\nabla_nY^{k\ell m}_{(1)\,ab}&=& \frac{5-k(k+2)}{a^2}Y^{k\ell m}_{(1)\,ab},
  \label{e:EigTensorY1}\\
  \nabla^n\nabla_nY^{k\ell m}_{(2)\,ab}&=& \frac{5-k(k+2)}{a^2}Y^{k\ell m}_{(2)\,ab},
  \label{e:EigTensorY2}\\
  \nabla^n\nabla_nY^{k\ell m}_{(3)\,ab}&=& \frac{6-k(k+2)}{a^2}Y^{k\ell m}_{(3)\,ab},
  \label{e:EigTensorY3}\\
  \nabla^n\nabla_nY^{k\ell m}_{(4)\,ab}&=& \frac{2-k(k+2)}{a^2}Y^{k\ell m}_{(4)\,ab},
  \label{e:EigTensorY4}\\
  \nabla^n\nabla_nY^{k\ell m}_{(5)\,ab}&=& \frac{2-k(k+2)}{a^2}Y^{k\ell m}_{(5)\,ab}.
  \label{e:EigTensorY5}
\eea
The divergences of the symmetric tensor harmonics are given by
\begin{eqnarray}
\nabla^aY^{k\ell m}_{(0)\,ab}&=& \frac{\sqrt{k(k+2)}}{\sqrt{3}a}
Y^{k\ell m}_{(0)\,b},\label{e:DivTensorY0}\\
\nabla^aY^{k\ell m}_{(1)\,ab}&=& 
-\frac{\sqrt{(k-1)(k+3)}}{\sqrt{2}a}Y^{k\ell m}_{(1)\,b},\label{e:DivTensorY1}\\
\nabla^aY^{k\ell m}_{(2)\,ab}&=& 
-\frac{\sqrt{(k-1)(k+3)}}{\sqrt{2}a}Y^{k\ell m}_{(2)\,b},\label{e:DivTensorY2}\\
\nabla^aY^{k\ell m}_{(3)\,ab}&=& -\frac{\sqrt{2(k-1)(k+3)}}{\sqrt{3}a}
Y^{k\ell m}_{(0)\,b},\label{e:DivTensorY3}\\
\nabla^aY^{k\ell m}_{(4)\,ab}&=& 0,\label{e:DivTensorY4}\\
\nabla^aY^{k\ell m}_{(5)\,ab}&=& 0\label{e:DivTensorY5},
\end{eqnarray}
and their traces are
\begin{eqnarray}
  g^{ab}Y^{k\ell m}_{(A)\,ab} &=&\left\{
  \begin{array}{c l}
    \sqrt{3}Y^{k\ell m}, & \qquad {\scriptstyle A}=0, \\
    0, & \qquad 1\le {\scriptstyle A} \le 5. \\
  \end{array}
  \right.\label{e:TraceIdentities}
\end{eqnarray}
Now one can expand an arbitrary symmetric tensor $s_{ab}$ on $\mathbb{S}^3$ as:
\begin{eqnarray}
  s_{ab}=\sum_{A k \ell m} \!\!
\tilde{s}^{k\ell m}_{(A)}
\,Y^{k\ell m}_{(A)\,ab}, \label{tensorExp}
\end{eqnarray}
where, as in the vector harmonic expansion, the range of allowed values for $k$ depends on $A$, the range of allowed values for $\ell$ depends on $k$, and the range of allowed values for $m$ depends on $\ell$, but again, the details are not important yet.
By inserting the orthonormality relations
\bea
  &&\delta_{(A)(B)}\delta^{\,kk'}\delta^{\,\ell\ell'}\delta^{\,mm'}=
\frac{1}{a^3}\int g^{ac}g^{bd}\,Y^{\,k\ell m}_{(A)\,ab}\,Y^{\,*k'\ell' m'}_{(B)\,cd}
\!\sqrt{g}\,
d^{\,3}x,\qquad
\label{e:TensorOrthoNormal}
\eea
into \eqref{tensorExp}, the expansion coefficients $\tilde{s}^{k\ell m}_{(A)}$ can be written as
\begin{eqnarray}
\tilde{s}^{k\ell m}_{(A)}=\frac{1}{a^3}\int g^{ac}g^{bd}
s_{ab} \,Y^{*k\ell m}_{(A)\,cd}\, \sqrt{g}\,d^{\,3}x.
\label{e:TensorCoefficients}
\end{eqnarray}
\\
\indent Now we are in a position to compute the expansion coefficients of $D_\Gamma X^i_a$. First, we identify the Lie-algebra valued one-forms $X^i_a$ and $D_\Gamma X^i_a$ with rank-two tensors $X_{ab} = e_{ib} X^i_a$, $D_\Gamma X_{ab} = e_{ib}D_\Gamma X^i_a$. 
Now we can calculate:
\bea
\tilde{D}_\Gamma\tilde{X}^{k\ell m}_{(0)} &=& \frac{1}{a^3}\int g^{ac}g^{bd} \epsilon_a{}^{\alpha\beta}\nabla_\alpha X_{\beta b}Y^{*\, \klm}_{(0)\, cd} \sqrt{g}\,d^3 x \\
&=& -\frac{1}{a^3}\int g^{ac}g^{bd}\epsilon_a{}^{\alpha\beta}X_{\beta b}\nabla_\alpha\!\left(\tfrac{1}{\sqrt{3}}Y^{* k\ell m} g_{cd}\right)\!\sqrt{g}\,d^3 x \\
&=& -\frac{1}{a^3}\int\frac{\sqrt{k(k+2)}}{\sqrt{3}a}\epsilon^{\alpha\beta b}X_{\beta b}Y^{* k\ell m}_{(0)\, \alpha}\!\sqrt{g}\, d^3 x \\
&=& -\frac{\sqrt{k(k+2)}}{\sqrt{3}a}\tilde{\alpha}^{k\ell m}_{(0)}.
\eea
\bea
\!\!\!\!\!\tilde{D}_\Gamma\tilde{X}^{k\ell m}_{(1)} &=& -\frac{1}{a^3}\int \epsilon^{c\alpha\beta}X_\beta{}^d\nabla_\alpha Y^{* k\ell m}_{(1)\, cd}\sqrt{g}\,d^3 x \nn \\
&=& -\frac{1}{a^3}\int \epsilon^{a\alpha\beta}X_\beta{}^b\nabla_\alpha \left(\frac{a}{\sqrt{2(k-1)(k+3)}}\left(\nabla_a Y^{* k\ell m}_{(1)\, b} + \nabla_b Y^{* k\ell m}_{(1)\, a}\right)\right)\!\sqrt{g}\,d^3 x \nn \\
&=& -\frac{1}{a^3}\int \frac{a}{\sqrt{2(k-1)(k+3)}} \epsilon^{a\alpha\beta}X_\beta{}^b\left( -\tfrac{1}{2}R_{\alpha a}{}^\lambda{}_b Y^{*\,k\ell m}_{(1)\,\lambda} + \nabla_b\nabla_\alpha Y^{*\, k\ell m}_{(1)\, a} - R_{\alpha b}{}^\lambda{}_aY^{*\, k\ell m}_{(1)\,\lambda} \right)\!\sqrt{g}\,d^3 x. \nn \\
\eea
Using $R_{ab}{}^c{}_d = \tfrac{1}{a^2}\left( \delta^c{}_a g_{bd} - \delta^c{}_b g_{ad} \right)$, the terms with curvature tensors above combine to give $\frac{2}{a\sqrt{2(k-1)(k+3)}}\tilde{\alpha}^{k\ell m}_{(1)}$, so we have 
\bea
\tilde{D}_\Gamma\tilde{X}^{k\ell m}_{(1)} = Q_{(1)}^{k\ell m} + \frac{2}{a\sqrt{2(k-1)(k+3)}}\,\tilde{\alpha}^{k\ell m}_{(1)}, \nn \\
\eea
where 
\bea
Q_{(1)}^{k\ell m} &=& -\frac{1}{a^3}\int \frac{a}{\sqrt{2(k-1)(k+3)}} \epsilon^{a\alpha\beta}X_\beta{}^b \nabla_b\nabla_\alpha Y^{*\, k\ell m}_{(1)\, a} \!\sqrt{g}\,d^3 x. \nn \\
&=& \frac{1}{a^3}\int\frac{a}{\sqrt{2(k-1)(k+3)}}X_\beta{}^b\nabla_b\left(\frac{k+1}{a}g^{\beta\lambda}Y^{*\, k\ell m}_{(2)\,\lambda}\right)\!\sqrt{g}\, d^3 x \nn \\
&=& \frac{1}{a^3}\int\frac{k+1}{\sqrt{2(k-1)(k+3)}}(s^{ab} + \alpha^{ab})\nabla_b Y^{*\, k\ell m}_{(2)\,a}\!\sqrt{g}\, d^3 x \nn \\
\eea
where in going from the first line to the second line we used \eqref{curl1}, and in the last line we have decomposed $X^{ab} = s^{ab} + \alpha^{ab}$, where $s^{ab}$ is the symmetric part (including the trace term) and $\alpha^{ab}$ is the antisymmtric part of $X^{ab}$. So now we have 
\be
Q_{(1)}^{k\ell m} = R_{(1)}^{k\ell m} + S_{(1)}^{k\ell m}
\ee
where
\bea
R_{(1)}^{k\ell m} &=& \frac{1}{a^3}\int\frac{k+1}{\sqrt{2(k-1)(k+3)}}s^{ab}\tfrac{1}{2}\left(\nabla_a Y^{*\, k\ell m}_{(2)\,b} + \nabla_b Y^{*\, k\ell m}_{(2)\,a}\right)\!\sqrt{g}\, d^3 x \nn \\
&=& \frac{1}{a^3}\int\frac{k+1}{2\sqrt{2(k-1)(k+3)}}s^{ab}\frac{\sqrt{2(k-1)(k+3)}}{a}Y^{*\,k\ell m}_{(2)\,ab}\!\sqrt{g}\,d^3 x \nn \\
&=& \frac{1}{a^3}\int\frac{k+1}{2a}\sigma^{ab}Y^{*\,k\ell m}_{(2)\,ab}\!\sqrt{g}\,d^3 x \nn \\
&=& \frac{k+1}{2a}\tilde{\sigma}^{k\ell m}_{(2)},
\eea
with $\sigma_{ab}$ being the trace-free symmetric part of $X_{ab}$, and
\bea
S_{(1)}^{k\ell m} &=& \frac{1}{a^3}\int\frac{k+1}{\sqrt{2(k-1)(k+3)}}\nabla_a\alpha^{ab}Y^{*\, k\ell m}_{(2)\,b} \!\sqrt{g}\, d^3 x \nn \\
 &=& \frac{1}{a^3}\int\frac{k+1}{\sqrt{2(k-1)(k+3)}}\nabla_a\left(\sum_{A' k' \ell' m'}\!\!\!\!\epsilon^{abc}\tilde{\alpha}^{k'\ell' m'}_{(A')}Y^{k'\ell' m'}_{(A')\, c}\right)Y^{*\, k\ell m}_{(2)\,b} \!\sqrt{g}\, d^3 x \nn \\
 &=& \frac{1}{a^3}\int\frac{k+1}{\sqrt{2(k-1)(k+3)}}\sum_{k' \ell' m'}\!\!\!\!\tilde{\alpha}^{k'\ell' m'}_{(1)}\left(-\frac{k+1}{a}g^{ab}Y^{k'\ell' m'}_{(2)\, a}\right)Y^{* \,k\ell m}_{(2)\, b} \!\sqrt{g}\, d^3 x \nn \\
 &=& -\sum_{k'\ell' m'}\frac{(k+1)^2}{a\sqrt{2(k-1)(k+3)}}\tilde{\alpha}^{k'\ell' m'}_{(1)}\frac{1}{a^3}\int g^{ab}Y^{k'\ell' m'}_{(2)\, a}Y^{*\, k\ell m}_{(2)\, b} \!\sqrt{g}\, d^3 x \nn \\
 &=& -\frac{(k+1)^2}{a\sqrt{2(k-1)(k+3)}}\tilde{\alpha}^{k\ell m}_{(1)},
\eea
where in going from the second line to the third line we used equations \eqref{curl0}, \eqref{curl1}, \eqref{curl2}, and the orthonormality condition \eqref{e:VectorOrthoNormal}, and in going from the fourth line to the last line we used the orthonormality condition \eqref{e:VectorOrthoNormal} again. So now we have 
\bea
Q_{(1)}^{k\ell m} &=& R_{(1)}^{k\ell m} + S_{(1)}^{k\ell m} \\
&=& \frac{k+1}{2a}\tilde{\sigma}^{k\ell m}_{(2)} -\frac{(k+1)^2}{a\sqrt{2(k-1)(k+3)}}\tilde{\alpha}^{k\ell m}_{(1)},
\eea
so that
\bea 
\tilde{D}_\Gamma\tilde{X}^{k\ell m}_{(1)} = \frac{k+1}{2a}\tilde{\sigma}^{k\ell m}_{(2)} +\frac{2 - (k+1)^2}{a\sqrt{2(k-1)(k+3)}}\tilde{\alpha}^{k\ell m}_{(1)}.
\eea 
The calculation for $\tilde{D}_\Gamma\tilde{X}^{k\ell m}_{(2)}$ proceeds in the same way as $\tilde{D}_\Gamma\tilde{X}^{k\ell m}_{(1)}$, so we will skip the details and record the result:
\be 
\tilde{D}_\Gamma\tilde{X}^{k\ell m}_{(2)} = \frac{2 - (k+1)^2}{a\sqrt{2(k-1)(k+3)}}\tilde{\alpha}^{k\ell m }_{(2)} + \frac{k+1}{2a}\tilde{\sigma}^{k\ell m}_{(1)}. 
\ee 
\bea
\!\!\!\!\!\!\tilde{D}_\Gamma\tilde{X}^{k\ell m}_{(3)} &=& -\frac{1}{a^3}\int\! \epsilon^{c\alpha\beta}X_\beta{}^d\nabla_\alpha Y^{*\, k\ell m}_{(3)\, cd}\sqrt{g}\,d^3 x \nn \\
&=& -\frac{1}{a^3}\int\! \frac{\sqrt{3}a}{\sqrt{2(k-1)(k+3)}}\epsilon^{c\alpha\beta}X_\beta{}^d\nabla_\alpha\left(\nabla_c Y^{*\, k\ell m}_{(0)\, d} + \frac{\sqrt{k(k+2)}}{\sqrt{3}a}Y^{*\,k\ell m}_{(0)\, cd}\right)\sqrt{g}\,d^3 x \nn \\
&=& -\frac{1}{a^3}\int\! \frac{\sqrt{3}a}{\sqrt{2(k-1)(k+3)}}\epsilon^{c\alpha\beta}X_\beta{}^d\left(-\tfrac{1}{2}R_{\alpha c}{}^\lambda{}_d Y^{*\,\klm}_{(0)\,\lambda} + \frac{k(k+2)}{3a^2}g_{cd} Y^{*\, \klm}_{(0)\, \alpha} \right)\sqrt{g}\, d^3 x \nn \\
&=& \ldots \nn \\
&=& -\frac{1}{a^3}\int\frac{-3 + k(k+2)}{\sqrt{3}a\sqrt{2(k-1)(k+3)}}\epsilon^{abc}X_{ab}Y^{*\,\klm}_{(0)\, c}\sqrt{g}\, d^3 x \nn \\
&=& \frac{\sqrt{(k-1)(k+3)}}{\sqrt{6}a}\tilde{\alpha}^\klm_{(0)},
\eea 
where in the omitted steps above we again used $R_{ab}{}^c{}_d = \frac{1}{a^2}(\delta^c{}_a g_{bd} - \delta^c{}_b g_{ad})$ and did some algebra.
\bea
\!\!\!\!\!\!\tilde{D}_\Gamma\tilde{X}^{k\ell m}_{(4)} &=& -\frac{1}{a^3}\int\epsilon^{c\alpha\beta}X_\beta{}^d\nabla_\alpha Y^{*\,\klm}_{(4)\, cd}\sqrt{g}\,d^3 x \nn \\
&=& \frac{1}{a^3}\int \left(\sigma^{ab}\frac{k+1}{a}Y^{*\,\klm}_{(5)\,ab} + \epsilon_\beta{}^{c\alpha}\nabla_\alpha \alpha^{\beta d}Y^{*\,\klm}_{(4)\, cd}\right)\sqrt{g}\, d^3 x \nn \\
&=& \frac{k+1}{a}\tilde{\sigma}^{\klm}_{(5)} + P_{(4)}^\klm,
\eea
where
\bea
P_{(4)}^\klm &=&  \frac{1}{a^3}\int\epsilon_\beta{}^{c\alpha}\nabla_\alpha \left( \sum_{A k' \ell' m'}\!\!\!\!\epsilon^{\beta d \lambda}\tilde{\alpha}^{k'\ell' m'}_{(A)} Y^{k'\ell' m'}_{(A)\, \lambda} \right)Y^{*\,\klm}_{(4)\, cd}\sqrt{g}\, d^3 x \nn \\
&=& \frac{1}{a^3}\int\left(g^{cd}g^{\alpha\lambda} - g^{c\lambda}g^{\alpha d}\right)\!\!\!\sum_{A k' \ell' m'}\!\!\!\!\tilde{\alpha}^{k'\ell' m'}_{(A)} \nabla_\alpha Y^{k'\ell' m'}_{(A)\, \lambda} Y^{*\, \klm}_{(4)\, cd}\sqrt{g}\, d^3 x \nn \\
&=& -\frac{1}{a^3}\int\sum_{A k' \ell' m'}\!\!\!\!\tilde{\alpha}^{k'\ell' m'}_{(A)}\nabla^d Y^{k'\ell' m'\, c}_{(A)}Y^{*\, \klm}_{(4)\, cd} \sqrt{g}d^3 x \nn \\
&\propto& -\frac{1}{a^3}\int\sum_{A = 0}^2\sum_{ k' \ell' m'}\!\!\!\! \tilde{\alpha}^{k'\ell' m'}_{(A)}Y^{k'\ell' m'\, cd}_{(Q(A))}Y^{*\, \klm}_{(4)\, cd}\sqrt{g}\, d^3 x = 0,
\eea
where in going from the second line to the third line we used the fact that $Y^{*\, \klm}_{(4)\, ab}$ is trace free and in the last line $Q(0) = 3$, $Q(1) = 1$ and $Q(2) = 2$. Finally, we see that $P^{\klm}_{(4)}$ vanishes by virtue of the orthonormality condition \eqref{e:TensorOrthoNormal}.
So we have
\be
\tilde{D}_\Gamma\tilde{X}^{k\ell m}_{(4)} = \frac{k+1}{a}\tilde{\sigma}^{\klm}_{(5)}.
\ee
\bea
\!\!\!\!\!\!\tilde{D}_\Gamma\tilde{X}^{k\ell m}_{(5)} &=& -\frac{1}{a^3}\int\epsilon^{c\alpha\beta}X_\beta{}^d\nabla_\alpha Y^{*\,\klm}_{(5)\, cd}\!\sqrt{g}\, d^3 x \nn \\
&=& -\frac{1}{a^3}\int\epsilon^{c\alpha\beta}X_\beta{}^d \frac{a}{2(k+1)}\nabla_\alpha\!\left( \epsilon_c{}^{\lambda\mu}\nabla_\lambda Y^{*\,\klm}_{(4)\, \mu d} + \epsilon_d{}^{\lambda\mu}\nabla_\lambda Y^{*\,\klm}_{(4)\, \mu c} \right)\! \sqrt{g}\, d^3 x \nn \\
&=& P^{\klm}_{(5)} + Q^{\klm}_{(5)},
\eea
where 
\bea
P^{\klm}_{(5)} &=& -\frac{1}{a^3}\int\frac{a}{2(k+1)}\left(g^{\alpha\lambda}g^{\beta\mu} - g^{\alpha\mu}g^{\beta\lambda}\right)X_\beta{}^d\nabla_\alpha\nabla_\lambda Y^{*\,\klm}_{(4)\,\mu d}\!\sqrt{g}\, d^3 x , \label{P5}\\
Q^{\klm}_{(5)} &=& -\frac{1}{a^3}\int\frac{a}{2(k+1)}\epsilon^{c\alpha\beta}\epsilon^{d\lambda\mu}X_{\beta d}\left( -R_{\alpha\lambda}{}^\rho{}_\mu Y^{*\,\klm}_{(4)\,\rho c} - R_{\alpha\lambda}{}^\rho{}_c Y^{*\,\klm}_{(4)\,\mu\rho} + \nabla_\lambda\nabla_\alpha Y^{*\,\klm}_{(4)\,\mu c} \right)\!\sqrt{g}\, d^3 x . \nn \\ \label{Q5}
\eea
Let's focus on $P^{\klm}_{(5)}$ first:
\bea 
P^{\klm}_{(5)} &=& -\frac{1}{a^3}\int\frac{a}{2(k+1)}\bigg[X^{\mu d}\frac{2 - k(k+2)}{a(k+1)}Y^{*\, \klm}_{(4)\, \mu d}  \nn \\ 
&& - X^{\lambda d}g^{\alpha \mu}\left(-R_{\alpha\lambda}{}^\rho{}_\mu Y^{*\, \klm}_{(4)\, \rho d} -R_{\alpha\lambda}{}^\rho{}_d Y^{*\, \klm}_{(4)\, \mu\rho} + \nabla_\lambda\nabla_\alpha Y^{*\, \klm}_{(4)\, \mu d} \right)\bigg] \nn \\
&=& \frac{k(k+2) - 2}{a(k+1)}\tilde{\sigma}^{\klm}_{(4)} + R^\klm_{(5)},
\eea 
where in the first line we used \eqref{e:EigTensorY4}, and where we have defined 
\bea
R^\klm_{(5)} &=& \frac{1}{a^3}\int\frac{a}{2(k+1)}X^{\lambda d}g^{\alpha\mu}\Big[-\tfrac{1}{a^2}\left(\delta^\rho{}_\alpha g_{\lambda\mu} - \delta^\rho{}_\lambda g_{\alpha\mu}\right)Y^{*\, \klm}_{(4)\, \rho d} \nn \\
&& -\tfrac{1}{a^2}\left(\delta^\rho{}_\alpha g_{\lambda d} - \delta^\rho{}_\lambda g_{\alpha d}\right)Y^{*\, \klm}_{(4)\, \mu\rho} + \nabla_\lambda\nabla_\alpha Y^{*\, \klm}_{(4)\,\mu d}\Big]\!\sqrt{g}\, d^3 x \nn \\
&=& \frac{1}{a^3}\int\frac{3}{2a(k+1)}X^{\lambda d}Y^{*\, \klm}_{(4)\,\lambda d}\sqrt{g}\, d^3 x \nn \\
&=& \frac{3}{2a(k+1)}\tilde{\sigma}^{\klm}_{(4)},
\eea 
where in going from the first equality to the second equality we used the trace-free and divergence-free properties of $Y^{*\, \klm}_{(4)\, ab}$ and did some algebra.
So we have 
\bea 
P^{\klm}_{(5)}
&=& \frac{k(k+2) - 2}{a(k+1)}\tilde{\sigma}^{\klm}_{(4)} + R^\klm_{(5)}, \nn \\
&=& \frac{k(k+2) - 2}{a(k+1)}\tilde{\sigma}^{\klm}_{(4)} + \frac{3}{2a(k+1)}\tilde{\sigma}^{\klm}_{(4)} \nn \\
&=& \frac{2k(k+2) - 1}{2a(k+1)}\tilde{\sigma}^{\klm}_{(4)}.
\eea 
Now let's turn to $Q^\klm_{(5)}$:
\be
Q^\klm_{(5)} = S^\klm_{(5)} + T^\klm_{(5)},
\ee 
where
\bea
\!\!\!\!S^\klm_{(5)} &=& \frac{1}{a^3}\int\epsilon^{c\alpha\beta}\epsilon^{d\lambda\mu}X_{\beta d}\frac{a}{2(k+1)}\bigg[\tfrac{1}{a^2}\left( \delta^\rho{}_\alpha g_{\lambda\mu} - \delta^\rho{}_\lambda g_{\alpha\mu} \right)Y^{*\, \klm}_{(4)\, \rho c} + \tfrac{1}{a^2}\left(\delta^\rho{}_\alpha g_{\lambda c} - \delta^\rho{}_\lambda g_{\alpha c}\right)Y^{*\, \klm}_{(4)\, \mu\rho}\bigg]\sqrt{g}\, d^3 x \nn \\
&=& \frac{1}{a^3}\int\frac{1}{2a(k+1)}X_{\beta d}\left( \epsilon^{c\beta}{}_\mu\epsilon^{d\rho\mu}Y^{*\, \klm}_{(4)\,\rho c} - \epsilon^{c\rho \beta}\epsilon_c{}^{d\mu}Y^{*\, \klm}_{(4)\,\mu\rho} \right) \sqrt{g}\, d^3 x \nn \\
&=& \frac{1}{a^3}\int\frac{1}{2a(k+1)}X_{\beta d}\left[\left(g^{cd}g^{\beta\rho} - g^{c\rho} g^{\beta d}\right)Y^{*\, \klm}_{(4)\, \rho c} - \left(g^{\rho d}g^{\beta\mu} - g^{\rho\mu} g^{\beta d}\right)Y^{*\, \klm}_{(4)\, \rho c} \right] \nn \\
&=& \frac{1}{a^3}\int\frac{1}{2a(k+1)}(X^{\rho c}Y^{*\, \klm}_{(4)\, \rho c} - X^{\mu\rho}Y^{*\, \klm}_{(4)\, \mu\rho })\sqrt{g}\, d^3 x = 0,
\eea
where in going from the first line to the second line we eliminated traces of the epsilon tensors, and in going from the third line to the fourth line we used the trace-free property of $Y^{*\, \klm}_{(4)\, ab}$. Now let's focus on $T^{\klm}_{(5)}$:
\bea 
T^{\klm}_{(5)} &=& -\frac{1}{a^3}\int\frac{a}{2(k+1)}\bigg[ \delta^\alpha{}_d\left(\delta^\beta{}_\lambda\delta^c{}_\mu - \delta^\beta{}_\mu\delta^c{}_\lambda\right) +  \delta^\alpha{}_\lambda\left(\delta^\beta{}_\mu\delta^c{}_d - \delta^\beta{}_d\delta^c{}_\mu\right) \nn \\ 
&& + \delta^\alpha{}_\mu\left(\delta^\beta{}_d\delta^c{}_\lambda - \delta^\beta{}_\lambda\delta^c{}_d\right)  \bigg]X_\alpha{}^d\nabla^\lambda\nabla_\beta\left(Y^{*\, \klm}_{(4)}\right)_c{}^\mu \sqrt{g}\, d^3 x \nn \\
&=& \frac{1}{a^3}\int\frac{a}{2(k+1)}X_\mu{}^d\left[\nabla^c\nabla_d\left(Y^{*\,\klm}_{(4)}\right)_c{}^\mu - \frac{2 - k(k+2)}{a}\left(Y^{*\, \klm}_{(4)}\right)_d{}^\mu\right]\sqrt{g}\, d^3 x \nn \\
&=& \frac{k(k+2) - 2}{2a(k+1)}\tilde{\sigma}^{\klm}_{(4)} + U^\klm_{(5)},
\eea 
where 
\bea 
U^\klm_{(5)} &=& \frac{1}{a^3}\int\frac{a}{2(k+1)}X_\mu{}^d\left(R_{cd}{}^c{}_\lambda Y^{*\, \klm \, \lambda\mu}_{(4)} + R_{cd}{}^\mu{}_\lambda Y^{*\, \klm \, c\lambda}_{(4)} \right)\sqrt{g}\, d^3 x \nn \\
&=& \frac{1}{a^3}\int\frac{a}{2(k+1)}X_\mu{}^d\left[\tfrac{2}{a^2}g_{d\lambda}Y^{*\, \klm \, \lambda\mu}_{(4)} + \tfrac{1}{a^2}\left(\delta^\mu{}_c g_{d\lambda} - \delta^\mu{}_d g_{c\lambda}\right)Y^{*\, \klm \, c\lambda}_{(4)} \right] \sqrt{g}\, d^3 x \nn \\
&=& \frac{1}{a^3}\int\frac{a}{2(k+1)}X_\mu{}^d\tfrac{3}{a^2}g_{d\lambda}Y^{*\, \klm \, \lambda\mu}_{(4)} \sqrt{g}\, d^3 x \nn \\
&=&\frac{3}{2a(k+1)}\tilde{\sigma}^\klm_{(4)}.
\eea 
So
\be 
Q^\klm_{(5)} = T^\klm_{(5)} = \frac{k(k+2) + 1}{2a(k+1)}\tilde{\sigma}^{\klm}_{(4)},
\ee 
and finally
\bea 
\tilde{D}_\Gamma\tilde{X}^{k\ell m}_{(5)} &=& P^\klm_{(5)} + Q^\klm_{(5)} \nn \\
&=& \frac{2k(k+2) - 1}{2a(k+1)}\tilde{\sigma}^{\klm}_{(4)} + \frac{k(k+2) + 1}{2a(k+1)}\tilde{\sigma}^{\klm}_{(4)} \nn \\
&=& \frac{3k(k+2)}{2a(k+1)}\tilde{\sigma}^\klm_{(4)}.
\eea 
We have now computed all of the expansion coefficients corresponding to the symmetric part of $D_\Gamma X_{ab}$. Next we need to compute the coefficients corresponding to the antisymmetric part. We will denote these coefficients by 
\bea
\tilde{D}_\Gamma\tilde{X}^\klm_{(A + 6)} &=& \frac{1}{a^3}\int D_\Gamma X_{ab}\epsilon^{abc}Y^{*\, \klm}_{(A)\, c}\sqrt{g}\, d^3 x \nn \\
&=& \frac{1}{a^3}\int\epsilon_a{}^{de}\nabla_d X_{eb}\epsilon^{abc}Y^{*\, \klm}_{(A)\, c}\sqrt{g}\, d^3 x \nn \\
&=& -\frac{1}{a^3}\int\left(g^{db} g^{ec} - g^{dc}g^{be}\right)X_{eb}\nabla_d Y^{*\,\klm}_{(A)\, c}\sqrt{g}\, d^3 x \nn \\
&=& -\frac{1}{a^3}\int \left(X^{ab}\nabla_b Y^{*\, \klm}_{(A)\, a} - g_{cd}X^{cd}\nabla^a Y^{*\, \klm}_{(A)\, a}\right)\sqrt{g}\, d^3 x \nn \\
&=& -\frac{1}{a^3}\int X^{ab}\nabla_b Y^{*\, \klm}_{(A)\, a}\sqrt{g}\, d^3 x - \delta^{(0)}{}_{\!\!(A)}\frac{\sqrt{3k(k+2)}}{a}\tilde{X}^\klm_{(0)} \nn \\
&=& \frac{1}{a^3}\int\nabla_b X^{ab}Y^{*\, \klm}_{(A)\, a}\sqrt{g}\,d^3 x - \delta^{(0)}{}_{\!\!(A)}\frac{\sqrt{3k(k+2)}}{a}\tilde{X}^\klm_{(0)},
\eea
where in going from the fourth line to the fifth line we used equations \eqref{e:DivVecY0}, \eqref{e:DivVecY1}, and \eqref{e:DivVecY2}. 
Now we can compute
\be 
\tilde{D}_\Gamma\tilde{X}^\klm_{(6)} = -\frac{\sqrt{3k(k+2)}}{a}\tilde{X}^\klm_{(0)} + P^\klm_{(6)} + Q^\klm_{(6)},
\ee
where 
\bea 
P^\klm_{(6)} &=& \frac{1}{a^3}\int \nabla_a \s^{ab} Y^{*\klm}_{(0)\,b}\sqrt{g}\, d^3 x , \nn \\
Q^\klm_{(6)} &=& -\frac{1}{a^3}\int \nabla_a \alpha^{ab} Y^{*\klm}_{(0)\,b}\sqrt{g}\, d^3 x.
\eea 
Let's focus on $P^\klm_{(6)}$ first:
\bea 
P^\klm_{(6)} &=& \frac{1}{a^3}\int \left(\sum_{A k' \ell' m'}\tilde{s}^{k'\ell' m'}_{(A)}\nabla_a Y^{k\ell' m'\, ab}_{(A)}\right)Y^{*\klm}_{(0)\,b}\sqrt{g}\, d^3 x \nn \\
&=& \frac{1}{a^3}\int \left[\sum_{k' \ell' m'}\left(\frac{\sqrt{k(k+2)}}{\sqrt{3}a}\tilde{X}^{k'\ell' m'}_{(0)} - \frac{\sqrt{2(k-1)(k+3)}}{\sqrt{3}a}\tilde{\sigma}^{k'\ell' m'}_{(3)} \right)Y^{k'\ell' m'\, b}_{(0)} + \perp \, \right]Y^{*\klm}_{(0)\,b}\sqrt{g}\, d^3 x \nn \\
&=& \frac{\sqrt{k(k+2)}}{\sqrt{3}a}\tilde{X}^{k\ell m}_{(0)} - \frac{\sqrt{2(k-1)(k+3)}}{\sqrt{3}a}\tilde{\sigma}^{k\ell m}_{(3)}.
\eea 
where $ ``\!\perp\!"$ denotes terms that vanish when integrated against $Y^{*\klm}_{(0)\,b}$, and we used equations \eqref{e:DivTensorY0}-\eqref{e:DivTensorY5} for the divergences of the symmetric tensor harmonics and the orthonormality relation \eqref{e:VectorOrthoNormal}.  For the other term we have
\bea 
Q^\klm_{(6)} &=& -\frac{1}{a^3}\int \left(\sum_{A k'\ell' m'}\tilde{\alpha}^{k'\ell' m'}_{(A)}\epsilon^{abc}\nabla_a Y^{k'\ell' m'}_{(A)\, c} \right)Y^{*\klm}_{(0)\,b}\sqrt{g}\, d^3 x \nn \\
&=& -\frac{1}{a^3}\int \sum_{A = 1}^2\sum_{k'\ell' m'}\tilde{\alpha}^{k'\ell' m'}_{(A)}\left(-\frac{k+1}{a}Y^{k'\ell' m'\, b}_{(Q(A))} \right)Y^{*\klm}_{(0)\,b}\sqrt{g}\, d^3 x = 0,
\eea 
where we used equations \eqref{curl0}, \eqref{curl1}, and \eqref{curl2}, and we defined $Q(1) = 2$, $Q(2) = 1$. Finally, we see that $Q^\klm_{(6)} = 0$, by virtue of the orthogonality relation \eqref{e:VectorOrthoNormal}. So in the end we have
\be 
\tilde{D}_\Gamma\tilde{X}^\klm_{(6)} = -\frac{2\sqrt{k(k+2)}}{\sqrt{3}a}\tilde{X}^\klm_{(0)} - \frac{\sqrt{2(k-1)(k+3)}}{\sqrt{3}a}\tilde{\sigma}^\klm_{(3)}.
\ee 
Using the same strategy, we can compute
\bea
\tilde{D}_\Gamma\tilde{X}^\klm_{(7)} &=& -\frac{\sqrt{(k-1)(k+3)}}{\sqrt{2}a}\tilde{\sigma}^\klm_{(1)} + \frac{1 + k(k+2)}{a(k+1)}\tilde{\alpha}^\klm_{(2)}, \\
\tilde{D}_\Gamma\tilde{X}^\klm_{(8)} &=& -\frac{\sqrt{(k-1)(k+3)}}{\sqrt{2}a}\tilde{\sigma}^\klm_{(2)} + \frac{1 + k(k+2)}{a(k+1)}\tilde{\alpha}^\klm_{(1)}.
\eea \
Finally, defining $\tilde{X}^{\klm}_{(A)} = \tilde\sigma^{\klm}_{(A)}$, $\tilde{X}^{\klm}_{(A + 6)} = \tilde\alpha^{\klm}_{(A)}$, we have
\begin{eqnarray}
\tilde{D}_\Gamma\tilde{X}^{k\ell m}_{(0)}
&=& -\frac{\sqrt{k(k+2)}}{\sqrt{3}a}\tilde{X}^{k\ell m}_{(6)}, \nonumber \\
\tilde{D}_\Gamma\tilde{X}^{k\ell m}_{(1)} &=& \frac{k+1}{2a}\tilde{X}^{k\ell m}_{(2)} +\frac{2 - (k+1)^2}{a\sqrt{2(k-1)(k+3)}}\tilde{X}^{k\ell m}_{(7)}, \nonumber \\
\tilde{D}_\Gamma\tilde{X}^{k\ell m}_{(2)} &=& \frac{k+1}{2a}\tilde{X}^{k\ell m}_{(1)} + \frac{2 - (k+1)^2}{a\sqrt{2(k-1)(k+3)}}\tilde{X}^{k\ell m }_{(8)} , \nonumber \\
\tilde{D}_\Gamma\tilde{X}^{k\ell m}_{(3)} &=& \frac{\sqrt{(k-1)(k+3)}}{\sqrt{6}a}\tilde{X}^\klm_{(6)}, \nn \\
\tilde{D}_\Gamma\tilde{X}^{k\ell m}_{(4)} &=& \frac{k+1}{a}\tilde{X}^{\klm}_{(5)}, \nn \\
\tilde{D}_\Gamma\tilde{X}^{k\ell m}_{(5)} &=& \frac{3k(k+2)}{2a(k+1)}\tilde{X}^\klm_{(4)}, \nn \\
\tilde{D}_\Gamma\tilde{X}^\klm_{(6)} &=& -\frac{2\sqrt{k(k+2)}}{\sqrt{3}a}\tilde{X}^\klm_{(0)} - \frac{\sqrt{2(k-1)(k+3)}}{\sqrt{3}a}\tilde{X}^\klm_{(3)}, \nn \\
\tilde{D}_\Gamma\tilde{X}^\klm_{(7)} &=& -\frac{\sqrt{(k-1)(k+3)}}{\sqrt{2}a}\tilde{X}^\klm_{(1)} + \frac{1 + k(k+2)}{a(k+1)}\tilde{X}^\klm_{(8)}, \nn \\
\tilde{D}_\Gamma\tilde{X}^\klm_{(8)} &=& -\frac{\sqrt{(k-1)(k+3)}}{\sqrt{2}a}\tilde{X}^\klm_{(2)} + \frac{1 + k(k+2)}{a(k+1)}\tilde{X}^\klm_{(7)}.
\end{eqnarray}

\bibliographystyle{ieeetr}
\bibliography{bib}

\end{document}